\documentstyle{article}
\author{Luis J. Boya
\footnote{e-mail address: luisjo@posta.unizar.es}
\and Antonio J. Segu\'\i-Santonja 
\footnote{e-mail address:segui@posta.unizar.es} \\
Departamento de F\'{\i}sica Te\'orica. Facultad de Ciencias.\\
Universidad de Zaragoza. 50009-Zaragoza, Spain}
\title{Topological charges and the genus of surfaces}
\begin{document}
\maketitle
\begin{abstract}
We show that the topological charge of the $n$-soliton solution of the
sine-Gordon equation $n[\phi]=[\int\partial_x \phi ]/2\pi$ is
related to the genus $g>1$ of a constant negative curvature compact 
surface described by this configuration. The relation is $n=2(g-1)$,
where $n=2 \nu$ is even. The moduli space of complex dimension
$B_g=3(g-1)$ corresponds precisely to the freedom to choosing the
configuration with $n$ solitons of arbitrary positions and velocities.
We speculate also that the odd soliton states will describe the
unoriented surfaces.
\end{abstract}

$1991$ MSC :$30$F, $32$G $15$, $35$Q $53$

\section{ Introduction }
\label{sec:I}
The sine-Gordon equation \cite{r1}, \cite{r2}
\begin{equation} \label{1-1}
\phi_{tt}-\phi_{xx}=-\sin\phi
\end{equation}
enjoys a great importance in physics:
\begin{enumerate}
\item In the lagrangian formalism it presents 
\emph{spontaneous breaking} of the discrete $Z$ symmetry
$\phi \to \phi+2{\pi}k$ and exhibits the attendant soliton
and multisoliton phenomenon. The single static
soliton is given as solution of the first order (Bogomolny)
equation; if $V(\phi)=W^2(\phi)/2$,
\begin{equation} \label{1-2}
{\phi'}{\equiv}{W(\phi)}=\sin{\phi}
\end{equation}
has as solution the profile
\begin{equation} \label{1-3}
\phi(x)=4 \arctan{\exp(x-x_0)}.
\end{equation}
\item The sine-Gordon equation has an \emph{auto-B\"acklund 
transformation} \cite{r3} which makes it possible to obtain the general
$n$-soliton solution \cite{r4}. The theory is exactly soluble also
by the Inverse Spectral Transform method \cite{r5}. The general 
solution contains multiple soliton/antisoliton configurations, as
well as breathers (soliton-antisoliton bound states) and background
\emph{noise}. The quantization is factible \cite{r6}, \cite{r7} and
indeed the formulas of the WKB approximation 
are already exact \cite{r8}.
\item The \emph{quantized} sine-Gordon theory is equivalent to the 
massive Thirring model \cite{r9} ; in fact, this duality, already
conjectured by Skyrme \cite{r10}, is the first case found of 
\emph{bosonization} of field theories with fermions. Explicit soliton
operators in terms of the $\phi$ field were also first exhibited here
by Mandelstam \cite{r11}, an example of the later much-studied 
\emph{vertex} operators. The quantum theory can also be made 
supersymmetric \cite{r12}.
\item The discrete $Z$ symmetry mentioned above can be seen as a
\emph{residual of the conformal invariance} of the (free) wave
equation in $1+1$ dimensions, namely
$\phi_{tt}-\phi_{xx}=0$, which is B\"acklund-transformable into the 
the conformal invariant Liouville equation
$\phi_{tt}-\phi_{xx}=e^{\phi}$ (see e.g. Olive \cite{r13}).
\item The sine-Gordon theory can be also seen as the \emph{non-lineal
sigma model} on the sphere 
$S^2$ \cite{r14}, \cite{r7}, and therefore it is
the simplest of the $\sigma$-models in $1+1$ which are exactly 
integrable, reflecting perhaps the fact that the naive $0+1$ 
"$\sigma$-model", the free motion on the sphere $S^n$, is 
superintegrable \cite{r15}.
\end{enumerate}

All this enhances the importance of the sine-Gordon system as a
toy model for some desirable properties of realistic theories, 
such as duality, bosonization, supersymmetry and softly-broken
conformal invariance \cite{r13}.

In this paper we focus our attention in the original motivation
for the appearance of the sine-Gordon equation, namely (\ref{1-1})
in light cone (characteristic) coordinates is precisely the equation
which describes the classical surfaces of constant negative 
curvature (Enneper, ca. $1880$). Because the compact representants
of these surfaces are topologically classified by the genus $g>1$,
and because also the manifold of solution of equation(\ref{1-1})
falls into classes labelled by the topological charge
\begin{equation}\label{1-4}
q(\phi)=n(\phi)=(1/{2\pi})\int{\phi_x dx}=(1/{2\pi})\{\phi(+\infty,t)
-\phi(-\infty,t)\},
\end{equation}
it is natural to relate the two topological invariants. The resulting 
relation is explained in section \ref{sec:III}
including the concordance of the \emph{moduli} space of this surfaces
under analytic transformations with the initial positions and
velocities of the $n$-soliton configuration. But first in section
\ref{sec:II}
we elaborate a bit on the geometry of the equation and its solutions.

\section{Surfaces of negative curvature}
\label{sec:II}
A surface $\Sigma$ embedded in ordinary space $R^3$ with 
\emph{gaussian} curvature \( K<0 \) everywhere has two asymptotic
directions in each point, separating the regions of positive and 
negative \emph{normal} curvature $\kappa$ 
( see e.g. Eisenhart \cite{r3}).
Taking coordinates $u,v$ parametrized by the arc-length, along these
directions, the metric becomes
\begin{equation}\label{2-1}
ds^2=du^2+2 F(u,v) du \hspace{0.1cm} dv+dv^2
\end{equation}
where $F$ is the cosine of the angle $\phi$ of parametric lines,
\begin{equation}\label{2-2}
F(u,v)=\cos \phi(u,v);
\end{equation}
all information on the surface is encoded in the function $F$.
It is easy to prove that these coordinates can be taken throughout
the surface; this is called a "Chebichev net" on the 
surface \cite{r16}.

The gaussian curvature is easily calculated,
\begin{equation}\label{2-3}
K=\frac{1}{1-F^{2}} \left(F_{uv} + \frac{F F_u F_v}{1-F^{2}} \right)
\end{equation}
or in terms of the $\phi$ angle
\begin{equation}\label{2-4}
\phi_{uv}=-K \sin \phi
\end{equation}
where $K$ is the gaussian curvature. If $K$ is a negative constant
(e.g. $-1/{a^2}$, say), this is of course the sine-Gordon equation
(\ref{1-1}) in light cone (or characteristic) 
coordinates, for $a=1$,
\begin{equation}\label{2-5}
u=\frac{t+x}{2} \qquad v=\frac{t-x}{2}.
\end{equation}

For this reason, equation (\ref{2-4}) was considered by Bianchi 
"l'equazione fondamentale di tutta la teoria delle superficie 
pseudosferique"(quoted by Coleman \cite{r6}).

On the other hand the \emph{compact oriented} 
surfaces of negative curvature,
which are Riemann surfaces, are perfectly well known and classified
(see e.g. \cite{r17}, \cite{r18}): there is 
the universal model, in the form
of a simply connected space ( with the topology of the plane), which
is usually presented in three forms \cite{r17}:
\begin{description}
\item[a)] The Minkowski model: the upper sheet $H^+$ of the 
two-sheeted hyperboloid in $R^3$ with the inherited metric from the
$++-$ metric in the ambient space $R^3$.
\item[b)] The Poincare disc $\Delta$, which is a stereographic
projection of the former from the vertex of the lower hyperboloid
\begin{equation}\label{2-6}
\Delta=\{z \in C \mid |z|<1 \} \qquad 
ds^2=\frac{4(dx^2+dy^2)}{(1-|z|^2)^2}.
\end{equation}
\item[c)] The upper half plane $U$ (Klein model)
\begin{equation}\label{2-7}
U=\{z\in C \mid \textrm{Im}z>0\} \qquad 
ds^2=\frac{dx^2+dy^2}{y^2}
\end{equation}
\end{description}
$H^+=\Delta=U$ are connected simply 
connected Riemann surfaces of constant
negative curvature ($-1$ by the given metric). 
For a detailed description
of these surfaces in relation to chaotic motion see \cite{r21}.

\emph{Any other} Riemann surface of the conformal class $K<0$
is obtained by quotienting by a 
subgroup $G$ of the modular group $M$,
which is a discrete automorphism group,
\begin{equation}\label{2-8}
\Sigma=U/G, \qquad G \subset M=PSL(2,Z).
\end{equation}

In fact, there are \emph{three} types of these surfaces:
\begin{enumerate}
\item The simply connected case, say $\Sigma=\Delta$ or $U$,
with the topology of the plane $C=R^2$; here $G=\{e\}$.
\item Those $\Sigma$ with $G=Z=\Pi_1 (\Sigma)$, topology of the
cylinder and conformally equivalent to \cite{r18} 
\begin{equation}\label{2-9}
\Delta^*=\Delta-\{0\} \qquad \textrm{or} \qquad 
\Delta_r=\{z \in C \mid 0<r<|z|<1\}
\end{equation}
\item All the other surfaces have a non-abelian, fundamental group
$\Pi_1$, are compact and topologically homeomorphic to a sphere with
$g$ handles (or holes), where $g>1$ ( $g=1$ correspond to the torus
$T^2$, which is of the conformal class flat). They can be represented
as union of tori $T=T^2$
\begin{equation}\label{2-10}
\Sigma_g=T\#T\#\cdots\#T \qquad (g\hspace{0.2cm} \textrm{times}, g>1)
\end{equation}
where $\#$ means the \emph{connected sum}, 
obtained by removing a little
open disc in each torus and soldering two of them by the boundary
circle \cite{r19}.
\item There remain only \emph{non-orientable} surfaces; the compact
ones are also classified by the genus $g$, and can be obtained
by the connected sum of \emph{projective planes} $RP^{2}=S^2/Z_2$
(antipodal map),
\begin {equation}\label{2-11}
{\Sigma'}_g=RP^2\#RP^2\#\cdots\#RP^2 \qquad 
(g\hspace{0.2cm} \textrm{times}, g>1).
\end {equation}
The case $g=1$ is the Klein bottle, of $K=0$ class.
\end{enumerate}

The \emph{homology} of these surfaces is easily computed, and 
it is \cite{r16}
\begin{eqnarray}\label{2-12}
\chi(\Sigma_g)&=&b_0-b_1+b_2=1-2g+1=2(1-g)
			\nonumber\\
\chi({\Sigma'}_g)&=&1-g+0=1-g
			\nonumber\\
H_1({\Sigma'}_g)&=&Z^g+Z_2
\end{eqnarray}

In all these surfaces we can chose a constant curvature metric,
by Riemann uniformization theorem \cite{r18}. However, as such
they cannot be embedded in $R^3$ with the induced metric from
$+++$(Hilbert theorem, \cite{r16}).

We shall need the area of the compact oriented surfaces which 
might be computed from the volume element
\begin{equation}\label{2-13}
dA=\sqrt{EG-F^2}\hspace{0.2cm} du \hspace{0.1cm} dv= 
|\sin \phi |\hspace{0.2cm} du\hspace{0.1cm} dv
\end{equation}
The area and the Euler number are connected through the fundamental
Gauss-Bonnet formula \cite{r16}
\begin{equation}\label{2-14}
\chi=\frac{1}{2\pi} \int K dA.
\end{equation}
This will be the key to identify the particular Riemann surfaces.

\section{Genus from soliton configurations}
\label{sec:III}
The solution of the sine-Gordon equation 
(\ref{1-1}) or (\ref{2-4}) are
classified by the topological charge(\ref{1-4}), namely

\begin{description}
\item[a)] $q=0$; the vacuum sector. It contains the vacuum solution
$\phi=2 \pi k$, the soliton -antisoliton scattering configurations, 
the soliton-antisoliton bound states (breather mode) and combinations
thereof.
\item[b)] $q=\pm 1$; it contains one soliton (resp. antisoliton) of
equation (\ref{1-3}), traslated and/or boosted, plus any solution of
a) above.
\item[c)] q arbitrary integer number; this is the multisoliton 
configuration; for example $q=+2$ will contain two solitons plus any
solution of a) above; etc.
\end{description}

Which negative curvature surfaces do these configurations belong to?
Let us start with the $q=2$ two soliton state. The connection with
the genus of the potential surface will be made through the
Gauss-Bonnet theorem: we can perform an area integration, applying
to (\ref{2-13}) and (\ref{2-14}) the equation (\ref{2-4})
$\sin \phi =\partial_{uv} \phi$:
\begin{equation} \label{3-1}
\int K dA=(-1)\int \mid \sin \phi \mid du\hspace{0.1cm} dv=
(-1) \int \partial_{uv} \phi\hspace{0.2cm} du\hspace{0.1cm} dv=
(-1) \phi\hspace{0.1cm} \textrm{(boundaries)}
\end{equation}

This is called "Hazzidaki's formula \cite{r16}. Now for the two-soliton
configuration it turns out that the algebraic sum of the boundary values
are just the jump in the $x=\pm \infty$ at $t=0$ minus the jump in
$t=\pm \infty$ at $x=0$, due to the relation between $u,v$ and $x,t$
(\ref{2-5}):
\begin{eqnarray}\label{3-2}
Area &= & \phi(x=+\infty,t=0) - \phi(x=-\infty,t=0)-{}
		\nonumber\\
& & {} \{\phi(x=0,t=+\infty)-\phi(x=0,t=-\infty)\}
\end{eqnarray}
which is just $4\pi$ for the two-soliton state (here there is not
jump in $t$ as $\phi(x=0,t)=\textrm{const.}$,by symmetry); the
explicit formula for the two-soliton solution in $x,t$ coordinates
is \cite{r1}, \cite{r4}
\begin{equation}\label{3-3}
\phi(x,t)=4 \arctan  \frac{-\cosh{\gamma v t} }{v \sinh{\gamma x} }
\end{equation}
where now $v$ is the relative velocity and $\gamma^2=(1-v^2)^{-1}$.

Therefore we are describing a $g=2$ surface, according the integral
formula
\begin{equation}\label{3-4}
\chi=\frac{1}{2 \pi} \int K dA = 2(1-g) = (-1)\frac{4 \pi}{2 \pi}
\qquad g=2.
\end{equation}

The generalization to $n=2 \nu$ \emph{even} number of solitons is 
immediate, because again there is no jump in $t$ while the jump 
in $x$ is given by $2 \pi n$
\begin{equation}\label{3-5}
2(1-g)=-\frac{(2 \pi)(2 \nu)}{2 \pi} \Longrightarrow g=\nu+1
\qquad (\nu>0)
\end{equation}
and it describes the compact oriented surface of genus $g$.
The span in $\phi$, namely $4 \pi \nu$, reflects the "holes" of
the surface. This is a satisfactory result. Of course, the integration
can be performed analytically also. The configuration with $2n$
antisolitons will presumably describe the same surface with changed 
orientation.

The correspondence goes along also with the integration parameters;
namely we can choose \emph{three} integration constants for each
soliton, the center, origin of time and velocity (the three parameters
of the $1+1$ Poincare group). But the \emph{moduli space} of the
surface of genus $g$ is known to be the Teichm\"uller space \cite{r17},
\cite{r18} of \emph{complex} dimension \cite{r20}
\begin{equation}\label{3-6}
B_g = \textrm{dim Teich} (\Sigma_g)= 3(g-1)= \frac{1}{2} \times 3
\times (2 \nu) = \frac{1}{2} (\#\textrm{real parameters})
\end{equation}
because $g=\nu-1$. So this is again in agreement.

We do not have a satisfactory answer for the odd-soliton case, for
which the integration is ill-defined. If we maintain (\ref{3-4}) for any
soliton number, i.e.
\begin{equation}\label{3-7}
-\chi=n\Longrightarrow g=n+1\hspace{0.2cm}  \textrm{for 
\emph{un}oriented surfaces},
\qquad \chi=1-g
\end{equation}
and we conjecture that this is true; in this case the soliton will
describe the "unoriented" pretzel, $g=2$. This goes on with the fact
that the soliton will be a fermion, and fermions are odd under 
full rotations. Again the concordance goes also with the moduli 
space, for which the freedom is now in \emph{real} dimension \cite{r17}
\begin{equation}\label{3-8}
B_g=3(g-1)=\#\textrm{param. of the n-sol. config.}
\end{equation}

There are still other surfaces (Cfr. section \ref{sec:II})
; we conjecture also that the breather mode, i.e. a non trivial
solution with q=0, will correspond to the non-compact case, e.g. to the
simply connected model $\Delta$ or U.

\section*{Acknowledgements} \nonumber
Discussions with M. Asorey and F. Falceto (Zaragoza) and M. Santander
(Va\-lla\-do\-lid) were useful. This work was supported in part by
research grants AEN96-1670 (CSIC) and ERBCHRX-CT92-0035.

\end{document}